\documentclass{aa}
\def\note #1]{{\bf #1]}}
\def\dd{\rm d}
\def\Fsurf{{\cal F}_{\rm surf}}
\def\Pas{P_{\rm as}}
\def\PLor{P_{\rm Lor}}
\def\xas{x^{\rm (as)}}
\def\muHz{\,\mu{\rm Hz}}
\def\delnuas{\delta^{\rm (as)}\nu}

\def\paper1{{Paper I}}
\def\figdir{.}
\def\vectora{{\mathchoice{\mbox{\boldmath$\displaystyle a$}}
{\mbox{\boldmath$\textstyle a$}}
{\mbox{\boldmath$\scriptstyle a$}}
{\mbox{\boldmath$\scriptscriptstyle a$}}}}
\usepackage{graphics}

\begin{document}

   \thesaurus{06     
              (06.15.1;  
	       06.09.1
)}

\title{Effects of line asymmetries on the determination of
solar internal structure}

\titlerunning{Effects of line asymmetries}
\authorrunning{Rabello-Soares et al.}
 
\author{M.C.~Rabello-Soares
          \inst{1}
	\and
	J.~Christensen-Dalsgaard
	   \inst{1,2}
	\and
	C.S.~Rosenthal
	   \inst{3}
        \and
	M.J.~Thompson
	   \inst{4}
	}

   \offprints{M.C.~Rabello-Soares}

   \institute{Teoretisk Astrofysik Center, Danmarks Grundforskningsfond
	      DK-8000 Aarhus C, Denmark \\
	      email: csoares@obs.aau.dk
	\and
	      Institut for Fysik og Astronomi, Aarhus Universitet,
	      DK-8000 Aarhus C, Denmark \\
	      email: jcd@obs.aau.dk
	\and
		Institute of Theoretical Astrophysics, University of Oslo,
		Blindern,
 Oslo N-0315, Norway \\
	      email: colin.rosenthal@astro.uio.no
	\and
		Astronomy Unit, Queen Mary and Westfield College,
                  London E1 4NS, UK \\
	      email: M.J.Thompson@qmw.ac.uk
	     }

   \date{Received / Accepted }

   \maketitle

   \begin{abstract}

Despite the strong evidence that the peaks in the spectrum of solar
oscillations are asymmetric, most determinations of mode frequencies
have been based on fits of symmetric Lorentzian profiles to the Fourier
or power spectra of oscillation time strings. The systematic errors
resulting from neglecting the line asymmetry could have serious effects
on inversions for the solar internal structure and rotation. Here we
analyse artificial data based on simple models of the intrinsic line asymmetry,
using GONG mode parameters with asymmetries found by one of us 
(Rosenthal \cite{rosenthal2}).
By fitting Lorentzians to the resulting spectra, we estimate the likely
properties of the errors introduced in the frequencies.
We discuss whether these frequency shifts have a form similar to
the near-surface layers uncertainties
and are therefore suppressed in the solar structure inversion.
We also estimate directly their contribution, if any, in the solar
sound-speed and density determinations using the SOLA technique.

      \keywords{sun: oscillations -- sun: interior}

   \end{abstract}

\section{Introduction}

There is strong evidence that the observed profiles of solar oscillation
are asymmetric (e.g. Duvall et al. \cite{duvall}).
This is thought to be a consequence of the localized nature of
the stochastic driving source, possibly combined with 
a contribution due to noise which is correlated with the driving 
(e.g. Gabriel \cite{gabriel}; Abrams \& Kumar \cite{abrams}; Roxburgh \& Vorontsov \cite{roxburgh};
Nigam et al. \cite{nigam3}; Rosenthal \cite{rosenthal1}b; Nigam \& Kosovichev 
\cite{nigam1}).
Yet in most analyses of helioseismic data the observed Fourier
or power spectra have been fitted with symmetric Lorentzian profiles. 
This leads to systematic errors in the inferred frequencies, of possibly
quite serious effect on the inversion for, e.g., the solar internal
structure.

Just how serious will be the effect on the inversion depends 
on the variation  of the frequency shift with 
mode frequency ($\nu$) and degree ($l$). Our aim then is first
to assess how the frequency shift due to fitting asymmetric 
peaks with Lorentzian profiles varies in the $l-\nu$ plane, and then 
to consider its effect on structural inversion, specifically a
SOLA inversion for sound speed and density.
This continues our earlier investigation of this problem
(Christensen-Dalsgaard et al. \cite{jcd2}, hereafter {\paper1}).

First we need a model of the line asymmetry. 
We use the following simple
representation of the asymmetric profiles in oscillation power
(see also Rosenthal \cite{rosenthal2}):
\begin{equation}
\Pas(\nu) = {\alpha_1^2 \over x^2 + \alpha_1^2} 
\left(1 - {2 x \over \alpha_4} \right) + {\alpha_1^2 \over \alpha_4^2} 
+N \; .
\label{eqn:pas}
\end{equation}
Here 
\begin{equation}
x = {\pi \over \Delta \nu} ( \nu - \nu_0) \; ,
\label{eqn:pas2}
\end{equation}
where $\Delta \nu$ is the frequency separation between modes of
adjacent radial orders.
Also, $\nu_0$ is the eigenfrequency of the mode,
and $\alpha_4$ is a measure of the asymmetry.
Simple algebra 
shows that, provided $\alpha_1^2\ll\alpha_4^2$, $P_{\rm as}(\nu)$ has a minimum 
value of $\simeq N$ at $x\simeq\alpha_4$. Thus $N$ is a measure of the ratio
of noise to signal in the power. 
In the limit $\alpha_4 \rightarrow \infty$ we recover
a Lorentzian profile with maximum value unity, and with
half width at half maximum (measured in terms of $x$) equal to $\alpha_1$,
sitting on a uniform background of level~$N$.

We note that
Nigam \& Kosovichev (\cite{nigam1}) presented a formally different,
but mathematically essentially equivalent, expression for the
asymmetrical line profile. This is characterized by the dimensionless
asymmetry parameter $B$ which, in our notation, is given by
$B = -\alpha_1 / \alpha_4$.

To estimate the systematic error in the frequency determination we have
carried out fits of Lorentzian profiles to timestrings of artificial data,
assumed stochastically excited and with a power envelope (see Section 2
for a more precise description) given by
Eq.~(\ref{eqn:pas}). This simulates the analysis of the actual data.
The fit results in a determination of the location
$\xas$ of that symmetric Lorentzian which best fits the
asymmetric power distribution. 
From this, the frequency error $\delnuas$ resulting from the
assumption of a Lorentzian profile is obtained as
\begin{equation}
\delnuas = {\xas \over \pi} \Delta \nu \; .
\end{equation}

Of course, $\xas$ depends on parameters $\alpha_1$ (linewidth),
$\alpha_4$ (asymmetry parameter) and $N$ (background). 
In order to assess how $\delnuas$ varies over the $l-\nu$ diagram,
we need to know how linewidth, asymmetry parameter and background  
vary with $l$ and $\nu$. We have used GONG data to estimate the
variation of these quantities.


Finally, we have used a SOLA inversion technique to calculate
the apparent
differences in sound speed and density related to the frequency error
$\delnuas$.

\section{Predicted shifts using artificial data}

We have considered
p-mode power spectra in the vicinity of a single mode.
These were constructed, on the assumption of stochastic excitation,
as normally distributed Fourier spectra
with zero mean and variance given by Eq.~(\ref{eqn:pas}).
To save time and still have good frequency resolution,
the simulated data frequency resolution was
taken to be
$\alpha_1$ divided by 5.

\begin{figure}[tb]
\resizebox{\hsize}{!}{\includegraphics{\figdir/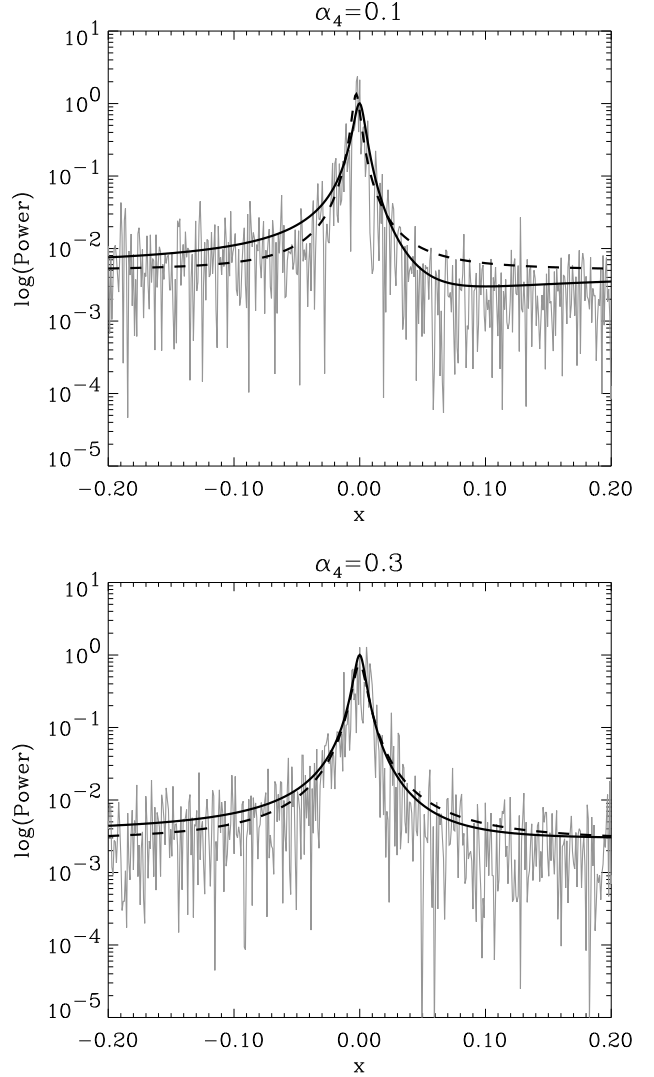}}
\caption
{Synthetic power spectra on a logarithmic scale generated using
$\alpha_1 = 0.0045$ and $N = 0.003$, 
as a function of $x$.
The two plots are two different realizations
using $\alpha_4 = 0.1$ and $0.3$.
The heavy continuous lines show the power envelopes $\Pas$
(cf. Eq.~\ref{eqn:pas}) used to generate the spectra,
and the dashed lines are the fitted Lorentzian profiles.
}
\label{fig_power}
\end{figure}

Each synthetic power spectrum ($P(\nu_i)$) was fitted by
minimizing the negative of the logarithmic likelihood function ($L$):
\begin{equation}
S(\vectora) = - \mbox{ln} (L) = \sum_i \left\{ \log[v(\vectora,\nu_i)] +
{P(\nu_i) \over v(\vectora,\nu_i)} \right\} \; ,
\end{equation}
where $v(\vectora,\nu_i)$ 
is the model of the 
variance of the spectrum, 
determined by the parameters $\vectora$.
The model was taken to be a symmetric Lorentzian profile,
\begin{equation}
v(\vectora,\nu) \ \equiv\ 
\PLor(\nu)\ =\ A {\gamma^2 \over (\nu - \nu_0)^2 + \gamma^2} + B \; ,
\label{eqn:lor}
\end{equation}
where $A$ is the amplitude, $\nu_0$ is the fitted frequency,
$\gamma$ is the half width at half maximum and $B$ is the background.
As in the fits of real data, the usual parameters were fitted:
the central frequency, and the logarithms of the
amplitude, linewidth and background noise level.
We have fitted over intervals of 100 times $\alpha_1$.
Fig.~\ref{fig_power} shows two
examples of `peaks' in such artificial power spectra,
and the discrepancy between the true line profile and the 
fitted Lorentzian.

\begin{figure}[tb]
\resizebox{\hsize}{!}{\includegraphics{\figdir/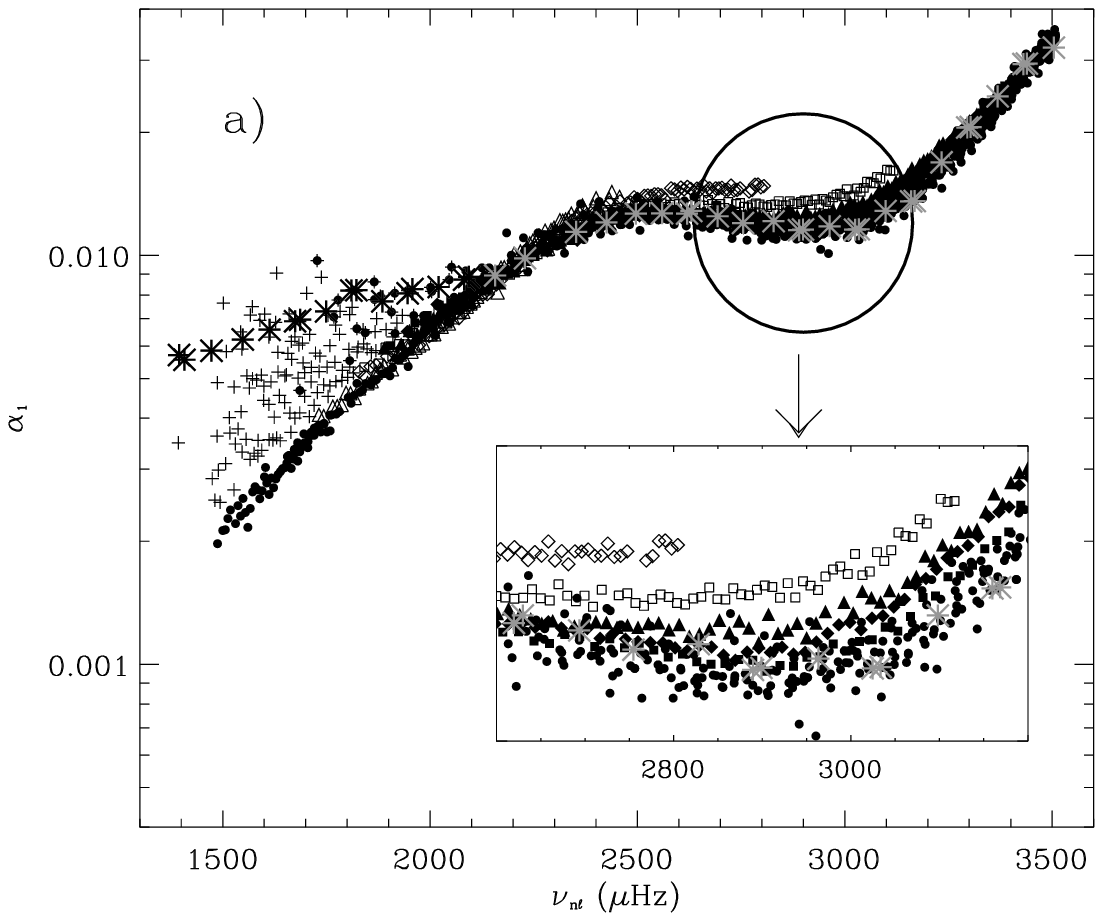}}
\resizebox{\hsize}{!}{\includegraphics{\figdir/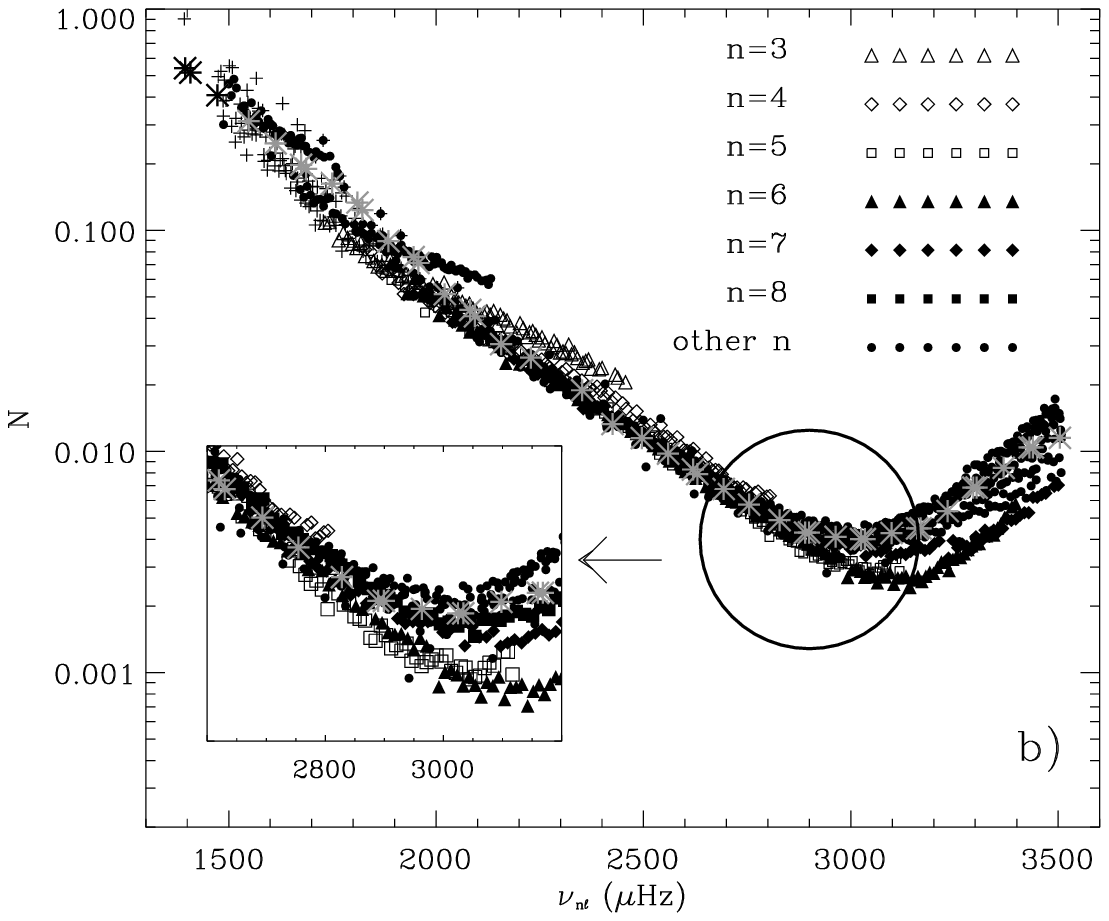}}
\caption
{
Parameters determining the frequency shifts.
Panel a) shows the dimensionless peak width $\alpha_1$
(measured in units of the large frequency separation $\Delta \nu$).
Panel b) shows the ratio $N$ between the background
noise and the peak power.
The stars are the extrapolated values for $l \leq 2$.
For later reference,
modes at low frequency with an $\alpha_1$ larger than
the general trend 
are indicated by crosses.
The inserts illustrate the dependence of the parameters
on mode order $n$, indicated by the symbol types listed in panel b).
}
\label{fig_nalpha1}
\end{figure}

We have used data obtained by the GONG project 
(e.g. Hill et al. \cite{hilletal})
to estimate
the parameters of Eq.~(\ref{eqn:pas})
used to generate the synthetic spectra corresponding to 
different points in the $l-\nu$ plane.
The data were determined in the GONG pipeline by
peak-bagging (using Lorentzian profiles) individual modes
in the averaged spectra
from six 3-month series of observations (GONG months 10-27).
For each $(n,l)$ we computed the $m$-averaged frequency, line width
(specified in the GONG tables as the FWHM), mode power and
background power; based on these averages, we determined the
parameters $\alpha_1$ and $N$
(assuming the limit $\alpha_4 \rightarrow \infty$)
as well as $\Delta \nu$
in Eqs~(\ref{eqn:pas}) and~(\ref{eqn:pas2}); the resulting $\alpha_1$ and $N$
are illustrated in Fig.~\ref{fig_nalpha1}.

\begin{figure}[t]
\resizebox{\hsize}{!}{\includegraphics{\figdir/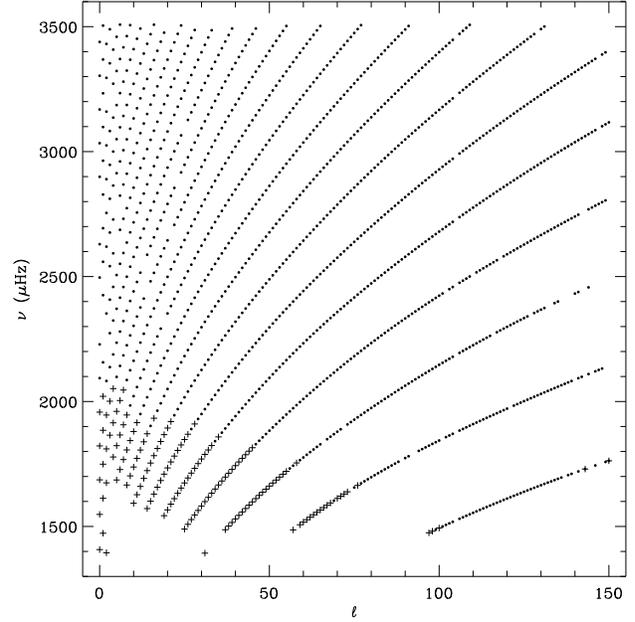}}
\caption[]
{
Multiplets for which GONG or BiSON (for $l \le 2$)
data have used in this investigation.
The crosses correspond to the crosses in Fig.~\ref{fig_nalpha1}.
}
\label{fig_lnu}
\end{figure}

Unfortunately, the GONG data contain 
few modes with $l=0$ and none with $l=1$ or 2.
To complement our mode set and make a 
more realistic representation of inversions to
infer solar structure, we have extrapolated the
parameters $\alpha_1$ and $N$ determined through
the GONG data for $l < 7$ modes to estimate those for $l < 3$.
The estimation was made for the data set observed by the BiSON
group and described by Basu et al. (\cite{basu2}).

We restrict the analysis to modes with frequency small\-er than
$3510 \muHz$; the mode set contains modes of degree up to 150
(Fig.~\ref{fig_lnu}).

The dimensionless asymmetry parameter $\alpha_4$ 
is determined by conditions near the upper turning point of the mode,
including the excitation and possibly the effects of correlated noise.
Although the details of these processes are as yet poorly
understood, it is perhaps not unreasonable to expect that they
depend little on degree, at least for the relatively modest
degrees considered here;
for in this case the horizontal scale of the modes far
exceeds the scales of the relevant convective processes.
Thus we might expect $\alpha_4$ to be a function of frequency alone.
We assume that this is precisely the case and estimate 
$\alpha_4$ from the dimensional results for low-degree modes in Fig.~4
of Rosenthal (\cite{rosenthal2}),
scaling with the value of $\Delta \nu$ for low-degree modes:
$\mbox{log}(\alpha_4) = -2.70+9.52 \times 10^{-4}~\nu$,
where $\nu$ is given in $\mu$Hz.
We note that this $\alpha_4$ and our estimate of $\alpha_1$ are
essentially consistent with the asymmetry parameter $B$
obtained from analysis of MDI (Toutain et al. \cite{toutain})
and GOLF (Thiery et al. 1999) data.

\begin{figure}[t]
\resizebox{\hsize}{!}{\includegraphics{\figdir/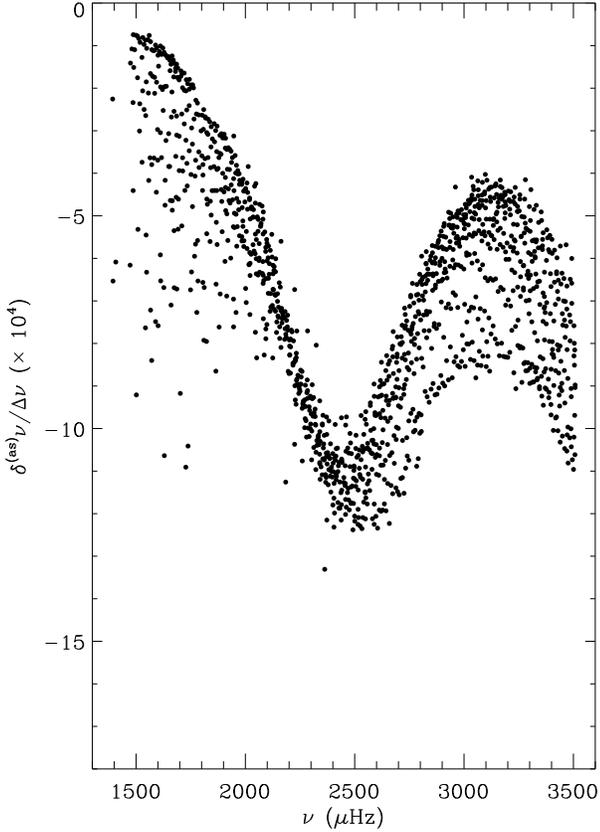}}
\caption[]
{
Dimensionless frequency shifts $\xas/\pi$ 
(i.e., $\delnuas$,
in units of the large frequency separation $\Delta \nu$)
obtained by fitting Lorentzians to the asymmetric profile
$\Pas(\nu)$ of Eq.~(\ref{eqn:pas}), 
using modal parameters based on GONG data. Each point 
corresponds to an average of 1000 simulations.
}
\label{fig_xas}
\end{figure}

We estimate the systematic frequency shifts by fitting symmetric
Lorentzians to asymmetric profiles, using the parameter values
estimated from the GONG data.
For each mode, 1000 simulations were performed;
averages of the estimated dimensionless frequency shifts 
$\xas/\pi$ are shown in Fig.~\ref{fig_xas}.
%

To a considerable extent, $\xas$ seems to be just a function of
frequency. 
This simple behaviour becomes more understandable if we consider
the dependence of the relevant parameters on $l$ and $\nu$.
As already stated, it is assumed that the dimensionless 
asymmetry parameter $\alpha_4$ is a function of frequency alone.
Fig.~\ref{fig_nalpha1} shows that the dimensionless line-width 
$\alpha_1$
and the noise-to-signal ratio $N$ for the GONG
mode set are predominantly functions of
frequency. 
(That this is so for $\alpha_1$ was indeed to be expected,
from the physical properties of the damping.)
However, they have also some small dependence on order $n$.

The dimensional frequency shifts $\delnuas$ will have
a much stronger $l$-dependence,
dominated by the dependence of $\Delta\nu$ on $l$.
However, it was shown in {\paper1} that $\Delta\nu$
is essentially inversely proportional to mode inertia 
(e.g. Christensen-Dalsgaard \& Berthomieu 1991);
this, therefore, is also the dominant $l$-dependence of the $\delnuas$.

\begin{figure}[t]
\resizebox{\hsize}{!}{\includegraphics{\figdir/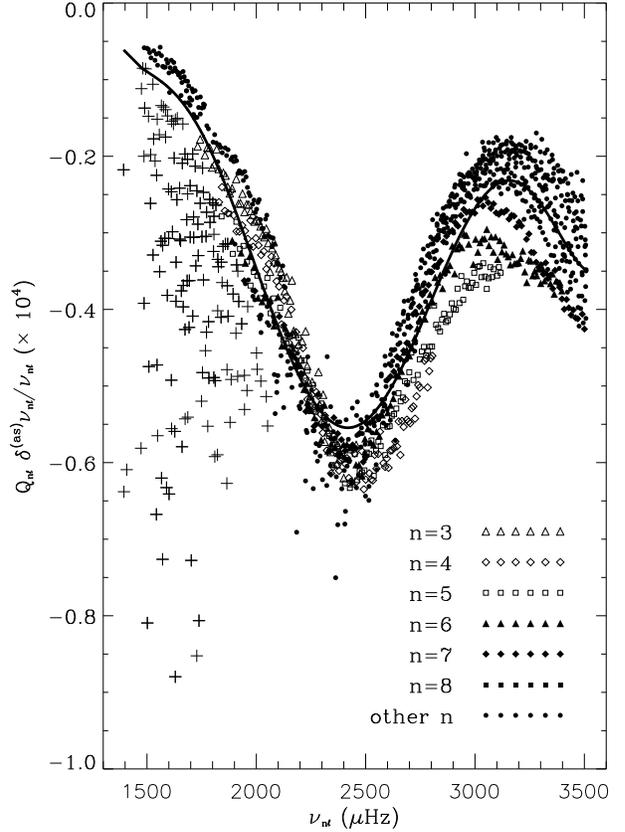}}
\caption[]
{
Relative frequency error resulting from the
assumption of a Lorentzian profile, multiplied by the
normalized mode inertia $Q_{nl}$.
The continuous curve is a fit to the points of
a polynomial of degree 6.
The crosses correspond to the crosses in Fig.~\ref{fig_nalpha1}.
}
\label{fig_relnu}
\end{figure}

\section{Effect on Inferred Solar Structure}

To test the effect on a solar structure inversion (e.g. for 
sound speed or density) of the systematic error from fitting asymmetric
peaks with Lorentzian profiles, we have run an inversion that
assumes the frequency differences between observation and a 
reference model to arise from differences in internal
structure, plus a possible contribution from the surface layers:
\begin{eqnarray}
{\delta \nu_{nl} \over \nu_{nl}} = 
\int K_{c^2,\rho}^{(nl)}(r) {\delta c^2\over c^2} (r) \ \dd r & + & \nonumber\\
 \int K_{\rho,c^2}^{(nl)}(r) {\delta \rho\over\rho}(r) \ \dd r & + &
\ {\Fsurf(\nu_{nl})\over Q_{nl}}\;. 
\label{eqn:inv}
\end{eqnarray}
We use a SOLA inversion which seeks to estimate the sound-speed (or density)
difference as a function of position within the Sun (see Basu et al. \cite{basu1}
for details). Since
the whole inversion process is linear, we do not need to create
a full set of artificial data: rather, we simply use
$\delnuas_{nl}/\nu_{nl}$ as our data, and this corresponds to the
systematic error that would be added to a real inversion of 
the GONG data 
due to fitting asymmetric peaks with Lorentzian profiles.

Note that our formulation in Eq.~(\ref{eqn:inv})
assumes that the difference
between the observed and model frequencies
in general contains a contribution of the form
$\Fsurf(\nu_{nl}) / Q_{nl}$
where $\Fsurf$, which as indicated is a function of frequency alone,
is determined by the near-surface errors in the model, 
and $Q_{nl}$ is the mode inertia normalized by the
inertia of a radial mode of the same frequency
(e.g. Christensen-Dalsgaard \& Berthomieu \cite{jcd_berthomieu}).
The SOLA inversion allows us to suppress the contribution from
the function $\Fsurf$ by imposing a constraint that would
completely remove such a function if it were
a polynomial of degree $\Lambda$ or smaller.
Typically we choose $\Lambda=6$: alternatively
we can simply choose not to impose this extra constraint at all.

\begin{figure}[t]
\resizebox{\hsize}{!}{\includegraphics{\figdir/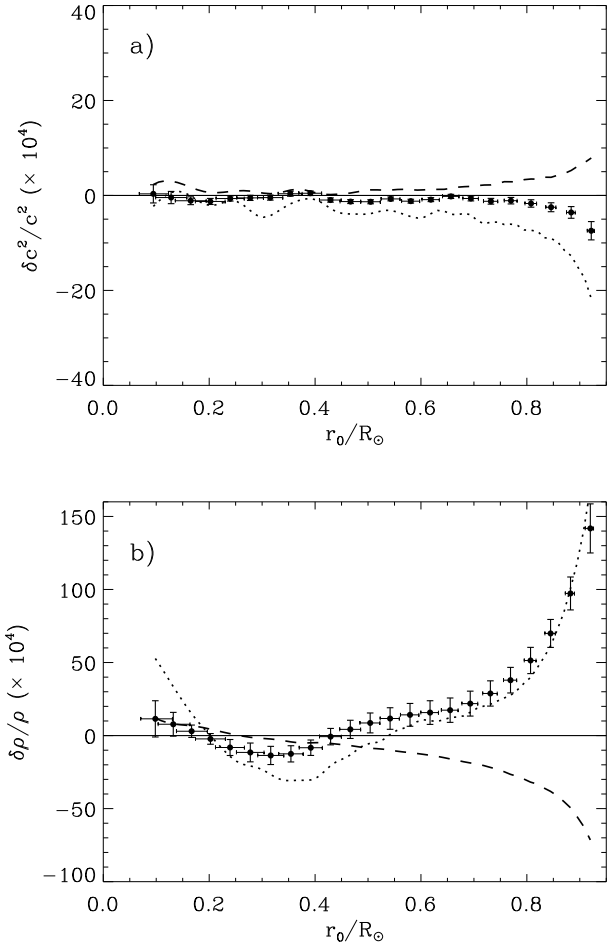}}
\caption[]
{
Relative difference in: (a) sound speed and (b) density
due to systematic errors in the inferred frequency resulting
from the assumption of a Lorentzian profile (Fig.~\ref{fig_relnu}).
The mode set for the inversions in each panel is the same;
the inversions assumed the errors of the observed data in the
combined GONG and BiSON data set.
The dotted line is for the average over
$2l+1$ estimations of $\delnuas$ (see below),
whereas 
the dashed line used the $\delnuas$ calculated with MDI linewidths
(see below).
}
\label{fig_inv_as}
\end{figure}

\begin{figure}[t]
\resizebox{\hsize}{!}{\includegraphics{\figdir/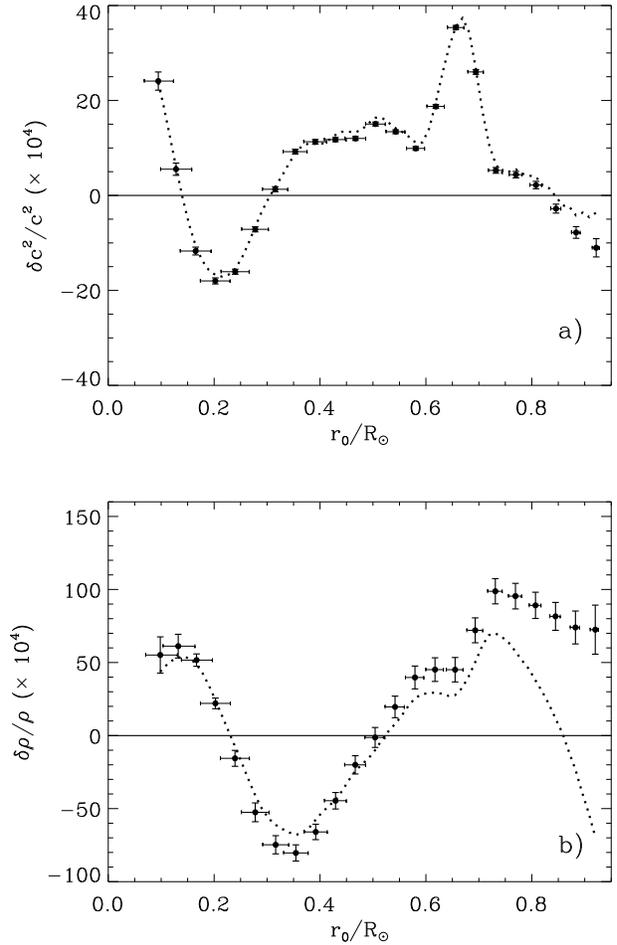}}
\caption[]
{
Relative difference in: (a) sound speed and (b) density
between Sun and the reference model using the observed frequencies calculated by
the GONG and BiSON projects (Lorentzian fitting).
The dotted line is the solution has been obtained by subtracting the
inferred relative differences (circles) in Fig.~\ref{fig_inv_as}, i.e.,
it shows the solution `corrected'
for the effect of fitting the data with a Lorentzian profile.
}
\label{fig_inv_sun}
\end{figure}

If the frequency shifts introduced
by mis-fitting the asymmetric peaks were also to be of the
form $\Fsurf(\nu_{nl}) / Q_{nl}$, 
they too would therefore be suppressed in the inversions.
In Fig.~\ref{fig_relnu}, we have plotted the relative frequency error
resulting from the assumption of a Lorentzian profile,
multiplied by the normalized mode inertia $Q_{nl}$.
As $Q_{nl}$ is asymptotically related to $\Delta \nu$,
Fig.~\ref{fig_relnu} is very similar to Fig.~\ref{fig_xas}.
The scattered
points at low frequency (crosses) are associated with the scatter in 
$\alpha_1$ (crosses in Fig.~\ref{fig_nalpha1}).
The linewidth determination at low frequency is particularly difficult 
because of its small magnitude. We have looked at other data sets, 
e.g. MDI data, and this scatter at low frequency is not present, 
which suggests that it is an artifact of a poor determination 
(see Fig.~\ref{fig_mdi}a, below).


Fig.~\ref{fig_relnu} shows 
that the relative frequency error due to fitting Lorentzians, when 
scaled by mode inertia, is in fact largely just a function of frequency;
thus one might expect
that much of it would be suppressed by the suppression of the 
surface term in the SOLA inversion.
The actual inferred sound-speed and
density differences from performing SOLA inversions of
$\delnuas_{nl}$ are illustrated in Fig.~\ref{fig_inv_as}.
For comparison, we show also in Fig.~\ref{fig_inv_sun}
the inferred sound-speed and density differences
between the Sun and Model~S of Christensen-Dalsgaard et al. (\cite{jcd1}),
using the $m$-averaged mode frequencies obtained by
the GONG project plus the BiSON mode frequencies for $l<3$
(described in Section~2).
For consistency with the results for the solar data,
in the inversions we assumed that the errors in the $\delnuas_{nl}$
were the same as the errors in these observed frequencies.
The important conclusion is that
the change in the inferred sound speed due to $\delnuas_{nl}$ is small
compared with the total sound-speed difference between the Sun and the model.
The same is true 
of the change in inferred density below the convection zone, for $r < 0.7R$.
However, it appears that the inference of density in the convection
zone is more sensitive to the effects of asymmetry, as described here.
The magnitude of the effect can perhaps be better appreciated in 
Fig.~\ref{fig_inv_sun}. There the dotted lines show the result
that would be obtained from inversion of solar frequencies
corrected for the effect of asymmetry,
i.e., from observed frequencies fitted
with an asymmetric profile given by Eq.~(\ref{eqn:pas}),
on the assumption that our representation of asymmetry is correct.
The only significant modification from the original solar inference is for
$\delta \rho/\rho$ at $r > 0.7R$: there we see that the inferred
density difference in the convection zone
is no longer nearly constant, in contrast with
our usual experience with density differences between pairs of solar
models (e.g. Christensen-Dalsgaard \cite{jcd3}).

We note that it actually makes rather little difference
in this case whether the SOLA inversion explicitly suppresses 
surface contributions or not:
the variation with frequency in Fig.~\ref{fig_relnu}
has little effect on the inversion even when the surface
constraint is not applied. 
In fact, even without taking explicitly account of a surface term,
the inversion provides
some suppression of contributions of this form:
this follows from the fact that
the near-surface effect of, e.g., $\delta c^2/c^2$
in Eq. (\ref{eqn:inv}) is also of a form similar to the surface term;
thus the localization implicit in the inversion leads to a partial
elimination of such contributions.
(Of course, the
surface term in frequency differences between the Sun and adiabatic
frequencies of a model is typically more than an
order of magnitude larger than
the differences in Fig.~\ref{fig_relnu},
so it is indeed important to apply
a strict surface constraint when inverting real data.)

\begin{figure}[t]
\resizebox{\hsize}{!}{\includegraphics{\figdir/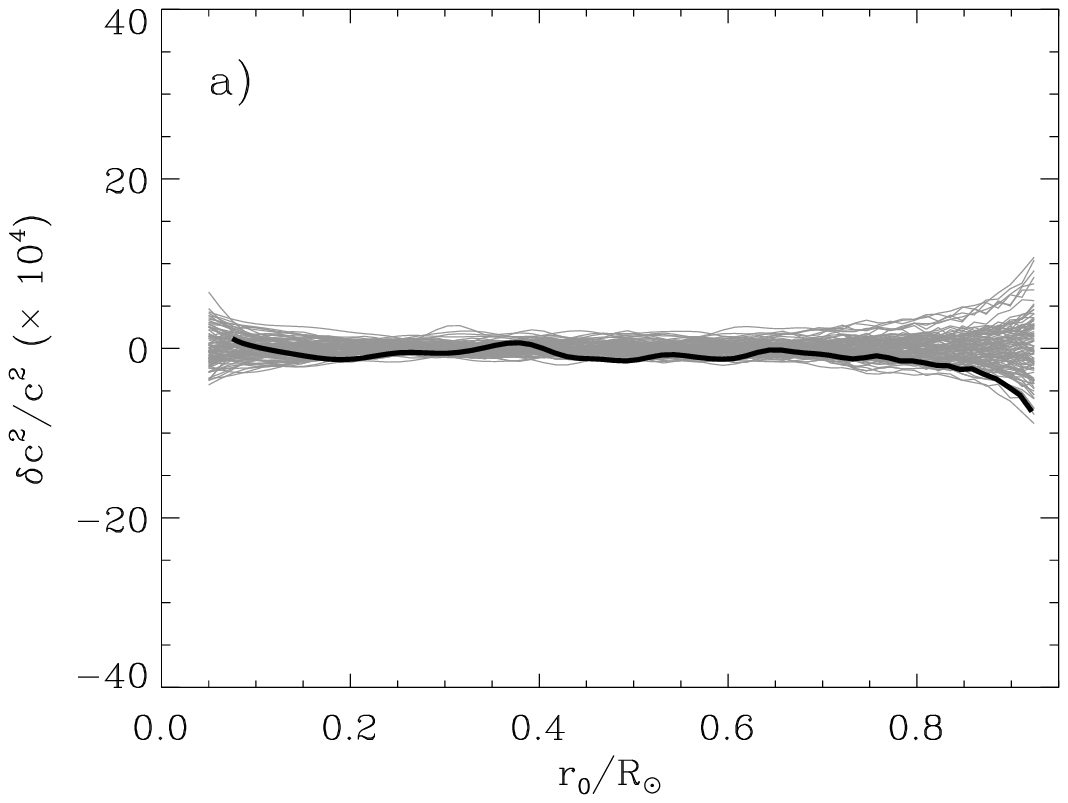}}
\resizebox{\hsize}{!}{\includegraphics{\figdir/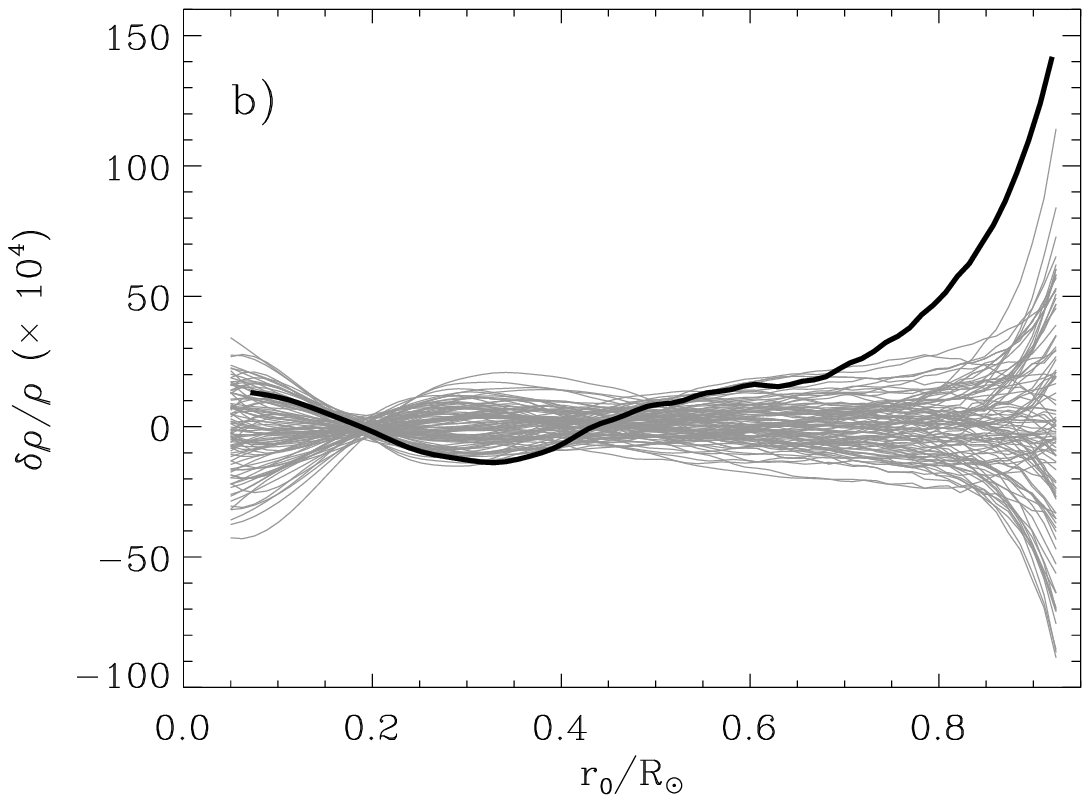}}
\caption[]
{Relative difference in: a) sound speed and b) density
due to 100 different noise realizations (see text). The thick line is
the relative difference due to $\delnuas$, same as the circles
in Fig.~\ref{fig_inv_as}.
}
\label{fig_noisy}
\end{figure}

It follows that the results of the inversions, shown in Fig.~\ref{fig_inv_as},
are dominated by the scatter of the points in Fig.~\ref{fig_relnu} around
the curve fitted as a function of frequency.
Much of this scatter could be due to noise
in the GONG parameter values. This then affects the inversions in much
the same way as random noise in the frequency data.
We have carried out inversions of
normally distributed random numbers with variances corresponding
to the spread in the points in Fig.~\ref{fig_relnu} from the fitted curve 
(neglecting the low-frequency values shown by crosses),
to demonstrate the effect of the scatter on our inversions.
The results are shown in Fig.~\ref{fig_noisy},
indicating that the sound-speed results are essentially consistent
with such a random distribution.
In addition, we have verified that the 
monotonic increase with $r$ in $\delta \rho/\rho$ in
the convection zone results from the scattered points at low
frequency, shown as crosses in Fig.~\ref{fig_relnu}.

\section{Discussion}

As described in Section~2,
when fitting the observations, the mode parameters were estimated for
each ($n,l,m$) and averaged over the $(2l+1)$ different
$m$ to give the values for
the multiplet ($n,l$). 
To reproduce the properties of the observations, 
it would therefore be more realistic 
to perform $2l+1$ simulations and average the estimated $\delnuas$, 
instead of averaging 1000 realizations as considered so far.
This will increase the scatter in the 
frequency shift, especially for low-degree modes.
The dotted lines in Fig.~\ref{fig_inv_as} show the
corresponding sound-speed and density differences.
The monotonic decrease in sound-speed and increase in density differences
towards the surface are still present,
somewhat enhanced for sound speed as a result of the larger
scatter in $\delnuas$. 
However, the variations are still generally small 
compared with the total difference between Sun and model 
(Fig.~\ref{fig_inv_sun}). 

\begin{figure}[t]
\resizebox{\hsize}{!}{\includegraphics{\figdir/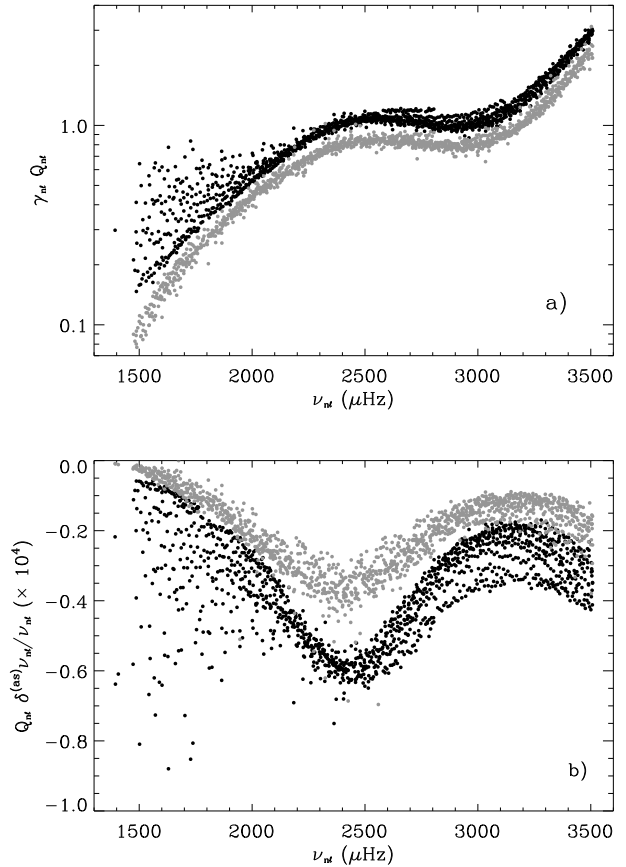}}
\caption[]
{
a) Normalized linewidth obtained from GONG (black dots) and
from MDI (gray dots) for a period corresponding to GONG months
29-31 by Schou (\cite{schou}).
b) Relative frequency error resulting from the
assumption of a Lorentzian profile multiplied by the
normalized mode inertia $Q_{nl}$ using linewidths obtained from:
GONG (black dots) and MDI (gray dots).
}
\label{fig_mdi}
\end{figure}

The inferred frequency shifts depend rather sensitively on the
parameters assumed for the modes, particularly their line widths,
and hence the use of parameter values based on just a single
data set might be cause for some concern.
We have already noted that noise in the GONG parameter determination
might have a substantial influence on the inversion results.
Another point is
that the GONG pipeline may systematically overestimate the linewidths.
Evidence for this assertion is shown in Fig.~\ref{fig_mdi}a, 
comparing the GONG linewidths used here with
MDI linewidths for a period corresponding to GONG months
29-31 obtained by the MDI Medium $l$ Program (Schou \cite{schou});
evidently, the MDI linewidths
are systematically smaller than those from GONG.
It seems probable that the systematic discrepancy arises from
the fact that the MDI parameter determination takes
the $m$-leakage into account, whereas the GONG determination did not.
(As an aside, we also note that
the scatter at low frequency in the GONG data -- crosses in
Fig.~\ref{fig_nalpha1}a -- is not present in the MDI data.) 

On the basis of this comparison it would obviously be
of interest to repeat our study using parameters obtained 
from the MDI pipeline.  Unfortunately,
there is not at the moment a good determination of the background
power when using the MDI pipeline 
(the fit is performed only in a very narrow window around each
peak, so that the background far from the peak cannot be determined
reliably).  However, we have tried using
MDI linewidths to estimate $\alpha_1$ and GONG data to estimate $N$: the
resulting $\delnuas/\nu$
are shown in Fig.~\ref{fig_mdi}b (gray dots) and compared with the
results we obtained with GONG linewidths (black dots). As one would
expect, the smaller MDI linewidths result in smaller values of $\delnuas$.
Thus, the results in the present study may be pessimistic in the
sense that if the GONG linewidths are overestimated, then this will
cause us also to overestimate the effects of line asymmetry.
Besides, the MDI linewidths do not show a dependence 
on degree (or order) at given frequency, 
in contrast to GONG data (cf. Fig.~\ref{fig_nalpha1}a).
Thus the dependence of $\delnuas$ on degree is small;
the degree dependence that remains
is probably introduced mostly by the $N$ parameter obtained
from GONG. The dashed lines in Fig.~\ref{fig_inv_as} show the
resulting inferred sound-speed and density differences.
The monotonic variation towards the surface is
substantially smaller than for the full GONG parameters,
particularly for density, and has the opposite sign;
this confirms our impression that the inferred structural 
variations in the solar interior (Fig.~\ref{fig_inv_as}, circles) 
are probably largely an artifact of the scatter in the 
GONG parameter determinations, which we have then used in our 
asymmetry model.

\section{Conclusions}

Our principal conclusion is that the frequency shift that
arises from
erroneously fitting the asymmetric line profiles with
symmetric Lorentz\-ians (as is commonly done at present) is rather benign:
it is predominantly of the same form as a structural near-surface
contribution -- viz. of the form $\Fsurf(\nu_{nl})/Q_{nl}$,
a function of frequency divided by mode inertia,
which is largely suppressed in the inversion,
and much of the departure from this behaviour is likely to be caused
by observational scatter in the GONG mode parameters which form the
basis for our study.
Thus, although 
there is some residual error introduced into the sound-speed and 
density inversions
at depth, its magnitude is generally very small. 
{\sl We therefore find no
evidence to suggest that ignoring line asymmetry has compromised the
helioseismic structural inversions published to date.}

One might worry that 
since the values of $N$ and $\alpha_1$ obtained from the GONG 
tables are themselves the result of a fit of a symmetric
Lorentzian profile, they are also subject to systematic error.
The symmetric fit picks out
the power level far from the peak as being the noise level. This
systematically over-estimates the true noise level which is given by the
power minimum or trough which lies close to the peak.
However, that error depends only on $\alpha_1$ and $\alpha_4$ and
should therefore also be a function only of frequency. 
On the other hand, the amplitude and line width
are only moderately affected by the asymmetry
(see {\paper1}).
Hence we do not expect this to affect our conclusions that $\xas$
is predominantly a function of frequency and that $\delnuas$ has
predominantly the same functional form as a near-surface contribution.

A recently published letter by Toutain et al. (\cite{toutain}) purports
to show a much more significant effect on the inferred
sound speed in the solar core from neglecting line asymmetry
when determining mode frequencies. Although we have not considered
precisely the same mode set that they did, we view the result of
Toutain et al. with some caution. They compared the inversion of
mode frequencies obtained by fitting observational data with 
symmetric Lorentzians with the inversion of mode frequencies in which
the low-degree modes only had been fitted with asymmetric profiles.
By fitting the low degrees asymmetrically, and the rest symmetrically,
it is quite probable that one will artificially introduce an
$l$-dependent error that does not scale like inverse mode mass and 
which will be erroneously interpreted in the inversion as
a spatial variation of the structure in the solar core.

\begin{figure}[t]
\resizebox{\hsize}{!}{\includegraphics{\figdir/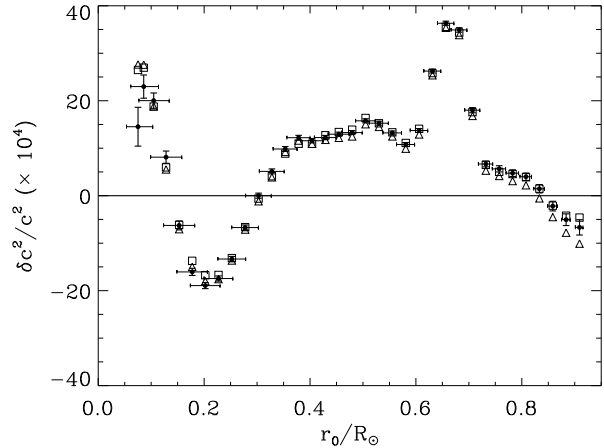}}
\caption[]
{
The circles with error bars show the results obtained with asymmetrically
fitted low-degree ($l \leq 2$) modes and symmetrically fitted otherwise ($l>2$).
The triangles represent the result of the symmetrical (Lorentzian)
fits to all modes (same as circles in Fig.~\ref{fig_inv_sun}a)
and the squares the asymmetrical fits to all modes (same as dotted line in
Fig.~\ref{fig_inv_sun}a).
To be compared with Fig.~5 in Toutain et al. (\cite{toutain}).
}
\label{fig_toutain}
\end{figure}

We did a similar experiment with our artificial data.
In Fig.~\ref{fig_toutain}, we compare the inversion of
mode frequencies obtained by fitting observational data with
symmetric Lorentzians (triangles) with the inversion of mode frequencies
which for the low-degree modes ($l \leq 2$) only have been 
corrected for the effect of asymmetry,
by applying our estimates of the frequency shift (circles with error bars).
Note that the inversion of the symmetrical fits (triangles) and
the inversion of
mode frequencies which have all been corrected
for the effect of asymmetry (squares)
agree quite well in the solar core,
as already shown in Fig.~\ref{fig_inv_sun}a. 
However, using the asymmetrical fits only for $l \leq 2$ modes introduces
a change in the solution in the solar core; this confirms our 
concern that such an inconsistent treatment of
asymmetry introduces an artificial $l$-dependence
and hence may affect the inversion in the core.
We note that the effect obtained here is more modest than that of
Toutain et al. (\cite{toutain}); the effect we find is quantitatively 
similar to what has been found by S.~Turck-Chi\`eze (private
communication). We have been able to make some qualitative 
comparisons with the asymmetry corrections applied by 
Toutain et al., thanks to T.~Toutain (private communication). It appears
that their asymmetry corrections for $l=1$ and $l=2$ are very similar and,
encouragingly, these agree with our simulated $\delnuas$ quite well;
but the $l=0$ corrections 
are rather different. 
Evidently such an effect would introduce some degree dependence 
and therefore radial
variation into Toutain et al.'s inversion results.


Although we conclude that ignoring line asymmetry has likely
not compromised helioseismic structural inversions to date,
we should like to emphasize the importance of
taking into account the asymmetric profile in the estimation of
mode parameters now and in the future.
With the SOHO satellite and the GONG network giving us longer time
series and with a better signal to noise ratio, 
the effect of asymmetry on the mode parameter determinations is
significant. 
This is particularly true as we shall be looking for ever more
subtle features in the solar interior.
Another important reason for fitting an asymmetric profile is 
the possibility of combining different observables.
For example, for solar oscillations observed in Doppler velocity and
continuum intensity, the asymmetry in their power spectra has
an opposite sign (Nigam \& Kosovichev \cite{nigam2})
which can lead to different estimates of the
mode parameters when fitting a symmetric profile. An inversion
of datasets with incompatible frequencies due to
different line asymmetries will be seriously compromised, because
the combined frequency error will not look like a single 
near-surface term and will in general be interpreted by the 
inversion method as a spatial variation inside the Sun. Likewise,
it is possible that non-contemporaneous data or
data from instruments with different levels of background noise 
would likewise contain systematic errors with serious consequences
for inversions, unless the peakbagging takes line asymmetry
into account.
We finally note that the determination of the asymmetry,
and other parameters of the modes, provides
crucial information about the excitation about the solar oscillations
(e.g. Rosenthal 1998a; Kumar \& Basu 1999; Nigam \& Kosovichev 1999).

\begin{acknowledgements}

We have utilized
data obtained by the Global Oscillation Network Group (GONG)
project, managed by the National Solar Observatory, a
Division of the National Optical Astronomy Observatories, which is
operated by AURA, Inc. under a cooperative agreement with the National
Science Foundation.
The data were acquired by instruments operated by the Big Bear Solar
Observatory, High Altitude Observatory, Learmonth Solar Observatory,
Udaipur Solar Observatory, Instituto de Astrofisica de Canarias, and
Cerro Tololo Interamerican Observatory.
We are most grateful to Rachel Howe for providing us with the `grand average'
peak-bagged data we have used in this investigation.
We thank Thierry Toutain for providing us with details of the MDI
low-degree asymmetry corrections, to inform our discussion in Section~5 
of the results of Toutain et al. (1998),
and Jesper Schou for the MDI mode linewidths.
The work was supported in part 
by the Danish National Research
Foundation through its establishment of the Theoretical Astrophysics Center,
by SOI/MDI NASA GRANT NAG5-3077,
and by the UK Particle Physics and Astronomy Research Council.
The National Center for Atmospheric Research is sponsored by the
National Science Foundation.

\end{acknowledgements}


\end{document}